**Singlet-triplet annihilation limits exciton yield in poly(3-hexylthiophene)**

Florian Steiner, Jan Vogelsang*, John M. Lupton

Institut für Experimentelle und Angewandte Physik, Universität Regensburg,

Universitätsstrasse 31, 93053 Regensburg, Germany

*jan.vogelsang@physik.uni-regensburg.de

Abstract

Control of chain length and morphology in combination with single-molecule spectroscopy techniques provide a comprehensive photophysical picture of excited-state losses in the prototypical conjugated polymer poly(3-hexylthiophene) (P3HT). A universal self-quenching mechanism is revealed, based on singlet-triplet exciton annihilation, which accounts for the dramatic loss in fluorescence quantum yield of a single P3HT chain between its solution (unfolded) and bulk-like (folded) state. Triplet excitons fundamentally limit the fluorescence of organic photovoltaic materials, which impacts on the conversion of singlet excitons to separated charge carriers, decreasing the efficiency of energy harvesting at high excitation densities. Interexcitonic interactions are so effective that a single P3HT chain of >100 kDa weight behaves like a two-level system, exhibiting perfect photon-antibunching.

Unraveling the photophysics of multichromophoric systems such as conjugated polymers (CPs) or light-harvesting complexes poses ongoing interdisciplinary challenges for chemists, physicists, material scientists and spectroscopists [1-3]. Some of the oldest yet still pressing questions in CP photophysics include the following: (i) What chemical unit absorbs and emits light in a CP? (ii) Which processes occur between absorption and emission events? And (iii) what is the interplay between efficient excitation energy transfer (EET) among chromophores and non-radiative fluorescence decay (quenching)? The last point is of particular relevance in organic photovoltaics: long-range EET is required to optimize charge photogeneration, yet promotes exciton quenching by dark states such as long-lived triplets or radicals. Answering these questions, and in particular elucidating which physical mechanisms are responsible for fluorescence self-quenching, is a prerequisite to fundamental understanding of CP photophysics. The heterogeneity of CPs with respect to their spectroscopic properties and morphological diversity has made single-molecule spectroscopy (SMS) an indispensable tool [4-9]. However, important parameters such as chain shape or size are difficult to assess in detail solely by SMS because these characteristics are convoluted with the photophysical observables themselves [10]. The formation, number and interaction pathways of chromophores in CPs correlate with changes in emission spectra and the degree of self-quenching [11-16]. However, the microscopic physical mechanism underlying this self-quenching phenomenon remains under debate [11, 13].

Here, we demonstrate that control of molecular weight (defined by the number-average molecular weight $M_n$) and chain morphology in combination with SMS techniques lead to a universal picture of self-quenching in CPs. As a model system, we use the prototypical CP poly(3-hexylthiophene) (P3HT, structure in Figure 2a) which is used widely in photovoltaics research [17]. In this material, the fluorescence quantum yield drops ~20-fold upon transitioning from well-dissolved (unfolded) P3HT chains in solution to aggregated (folded) chains in a bulk film [18]. By controlling chain morphology using solvent or matrix polarity in well-defined sub-populations of different $M_n$, fractionated by gel-permeation chromatography (GPC), we correlate microscopic photophysical observables with each other and with $M_n$. The increase in fluorescence self-quenching with increasing $M_n$ and morphological ordering can be rationalized by improved light-harvesting and singlet-triplet quenching. In other words, triplet excitons fundamentally limit the lifetime of singlet excitons, which can reduce the efficiency of organic solar cells.

The microscopic photophysical model of exciton self-quenching is formulated in Figure 1. Panel (a) shows the expected photoluminescence (PL) intensity versus $M_n$ for the two extreme morphologies: (i) unfolded, as it occurs for well-dissolved CPs in solution with few, if any, intrachain contacts [19]; and (ii) folded, as it arises for aggregated CPs in a bulk film with strong intrachain contacts [20, 21]. In case (i), the PL intensity should increase linearly with increasing $M_n$ because intrachain interchromophoric interactions are negligible. However, in case (ii), saturation is expected for folded chains provided that EET to one single emitting acceptor chromophore (marked red) occurs. This acceptor ultimately enters a triplet state so that singlet-triplet annihilation (STA) can quench the excitation energy of the complete CP chain. Panel (b) shows a simplified level scheme of the acceptor chromophore, which is excited at rate $k_{exc}$ from the singlet ground state $S_0$ to the first excited state $S_1$. $S_1$ decays (radiatively and non-radiatively) to $S_0$ at rate $k_{PL}$, or undergoes intersystem crossing into the triplet $T_1$ (rate $k_{isc}$). The acceptor chromophore emits photons, i.e. it is in an "on"-state for a characteristic duration $\tau_{on}$ as it cycles between $S_0$ and $S_1$. Following intersystem crossing, the chromophore enters an "off"-state for $\tau_{off}$, as illustrated in panel (b). Assuming that $k_{exc} \ll k_{PL} + k_{isc}$, $\tau_{on}$ can be described as follows [22]:

$$\tau_{on} = (k_{PL} + k_{isc})/(k_{isc} \cdot k_{exc}) \propto k_{exc}^{-1} \tag{1}$$

where $k_{exc}$ is proportional to the excitation intensity, $I_{exc}$, and the absorption cross-section. We propose that the entire chain [green, Fig.1(a)] absorbs light and transfers energy to the acceptor chromophore [red, Fig.1(a)]. Therefore, $k_{exc}$ is, to first-order approximation, directly proportional to $M_n$; Eq. (1) becomes

$$\tau_{on} \propto (M_n \cdot I_{exc})^{-1} \tag{2}$$

By STA, which gives rise to quenching of the complete CP once the acceptor chromophore enters the $T_1$-state, $\tau_{off}$ becomes independent of $M_n$:

$$\tau_{off} = (k_T)^{-1} \tag{3}$$

where $k_T$ is the reverse intersystem crossing rate (the inverse triplet lifetime). The average PL intensity of the single chain [marked red in Fig. 1(c)] is therefore given by

$$\langle PL \rangle \propto \frac{\tau_{on} \cdot PL_{max}}{\tau_{on} + \tau_{off}} \propto M_n \cdot I_{exc} \cdot \left(1 + \frac{M_n \cdot I_{exc}}{k_T}\right)^{-1} \quad (4)$$

where $PL_{max}$ denotes the maximum PL intensity during $\tau_{on}$, which is proportional to $k_{exc}$ and thus to $M_n$ and $I_{exc}$. Eq. (4) demonstrates the dependence of $\langle PL \rangle$: it saturates either with $I_{exc}$ or $M_n$ as sketched in the bottom of Figure 1(a), provided only one single accepting chromophore is present within the polymer.

To test this simple model, we revisited the dependence of $\langle PL \rangle$ on $M_n$ and $I_{exc}$ for unfolded and folded P3HT chains. P3HT was fractionated by GPC into 6 different $M_n$ samples with low polydispersity index (PDI), ranging from 19 kDa to 110 kDa versus polystyrene standards (see Table S1). We stress that the actual $M_n$ of P3HT may differ significantly by a constant factor from the values reported here due to the different rigidity of P3HT compared to the polystyrene standards. This difference is irrelevant for our study since the power exponent dependence of $M_n$ describing the radius of gyration stays the same, thus not affecting *relative* $M_n$ comparisons [23].

Perfectly unfolded isolated chains can only be arrived at in unpolar solution. Therefore, each sample was diluted to concentrations of $10^{-10} - 10^{-11}$ M in toluene, and $\langle PL \rangle$ was obtained by fluorescence correlation spectroscopy (FCS, see Fig. S1 and SI for details) [11]. Figure 2(a) shows single-chain $\langle PL \rangle$ in dependence of $M_n$. The PL intensity increases linearly as expected for well-dissolved chains in solution. This increase is a necessary control to validate that the $M_n$ values, obtained from GPC, are not compromised due to interchain aggregation at higher $M_n$ [11, 19]. The average PL intensity for *folded* chains cannot be measured in solution: inducing self-aggregation by raising solvent polarity will drive isolated single chains together, forming larger aggregates [24]. Instead, single chains are embedded in a polar host matrix, PMMA, where most chains fold to form highly ordered structures (i.e. self-aggregate) but remain isolated since they are "frozen" in the matrix [25]. With the molecules stationary, we cannot employ FCS which is a diffusion-based technique. Instead, $\langle PL \rangle$ of single P3HT chains was acquired by confocal scanning of several areas of $20 \times 20$ µm² (see Fig. S2 and SI for details on experimental apparatus). Diffraction-limited PL spots of the single chains were identified by an automated spot-finding algorithm. For each $M_n$, the average PL intensity was acquired over ~200 chains.

The average PL intensity of folded P3HT chains, plotted in Figure 2(b), saturates quickly with increasing $M_n$ and follows Eq. (4) [black curve in Fig. 2(b)]. The same saturation behavior is observed by raising $I_{exc}$ (Fig. S3). It is important to note that this observation precludes electronic aggregation (i.e. H-aggregation [28]) as the source of PL quenching in isolated chains since in this case the PL dependency would still have to be linear. The quenching of $\langle PL \rangle$ with $M_n$ is remarkable: as chain length increases, the brightness does not, implying that the self-quenching efficiency rises with chain length. Lin *et al.* previously attributed such self-quenching to the formation of unspecified "dark matter", in which parts of the folded chain do not participate in energy transfer or emission and are practically "invisible" [11]. In such a framework, excitation energy generated by absorption is not completely funnelled to an emitting chromophore [11]. Consequently, $\tau_{on}$, as defined by Eq. (1), would not depend on $M_n$. This situation can be tested experimentally.

Before proceeding to test Eq. (2) and (3) by measuring $\tau_{on}$ and $\tau_{off}$ on *folded* chains, we quantify the number of independently emitting chromophores, $N$, with respect to $M_n$ by examining the statistics of fluorescence photons, summarized in Figure 3 [26]. Strong photon antibunching can occur in P3HT [25]: excitons are funneled to the acceptor chromophore by EET and quenched by singlet-singlet annihilation, so that only one exciton emits at a time [27]. The samples were studied with a scanning confocal fluorescence microscope with two photodetectors in an interferometric Hanbury-Brown-Twiss geometry, under dry nitrogen atmosphere. PL transients from single folded chains were recorded and the probability of a chain emitting more than one photon at once was measured by quantifying photon coincidence on the two detectors. The number of independently emitting chromophores, $N$, was extracted from this probability (see Fig. S4 and Ref. [26] for details). The red dots in Figure 3 show $N$ in dependence of $M_n$ for folded P3HT chains. $N$ remains close to unity for folded chains over the $M_n$ range studied: on average, only one emitting chromophore (acceptor) is active at a given time. The nature of this acceptor chromophore remains subject to debate [28, 29], but is not strictly crucial to the present analysis. It was recently demonstrated that *intrachain* photophysics dominates the spectral changes observed when transitioning from the solution to the bulk [25], rather than interchain interactions as previously surmised [28]. While H-aggregation [28] may play a role in some features of the *bulk* photophysics, the dramatic red shift between solution and film (i.e. between solvated and collapsed chains) is primarily intrachromophoric in nature [25]. The relevant question is therefore what $M_n$ is necessary to form an acceptor chromophore in a folded morphology? We obtained ~100 single-molecule spectra for each $M_n$ sample for folded chains (Fig. S5). The black dots in Figure 3 show the average 0-0 transition energy, $E_{0-0}$, with respect to $M_n$ for folded chains. The transition energies lie at approximately 1.95 eV and shift to higher energies for the two lowest-$M_n$ samples. The spectral measurements indicate that the universal acceptor site is not always fully developed for $M_n \leq 45$ kDa [dashed line in Fig. 3], even though the shorter chains show photon antibunching.

Both data sets, photon statistics and PL peaks, provide a weight range (55–110 kDa) over which $\tau_{on}$ and $\tau_{off}$ of the acceptor can, in principle, satisfy Eq. (2) and (3). For $M_n \leq 45$ kDa, the acceptor site is not always formed in the intrachain aggregate, possibly because chain folding is not complete. The EET efficiency is therefore reduced, direct excitation of the acceptor is enhanced, and even the chromophoric triplet rates $k_{isc}$ and $k_T$ may be different, altering $\tau_{on}$ and $\tau_{off}$ values. On the other hand, if more than one acceptor is formed at higher $M_n$, excitation energy may be distributed sequentially between different chromophores within the molecular aggregate, increasing the apparent $\tau_{on}$ for each acceptor while maintaining $N=1$.

We proceed by measuring $\tau_{on}$ and $\tau_{off}$ in dependence of $M_n$. PL intensity transients of single folded chains, exemplified in Figure 4(a), were analyzed by passing the fluorescence through a 50:50 beam splitter and computing a second-order cross-correlation of the two detector signals [30]. Subsequent fitting of the cross-correlation, shown in Figure 4(b), by a three-state (exponential) model allows the extraction of $\tau_{on}$ and $\tau_{off}$ [see Fig. S6 and SI for details] [31]. Only transients exhibiting single-step photobleaching were considered (~90% of all single chains). Figure 4(c) displays average $\tau_{on}$ and $\tau_{off}$ values from ~100 PL transients obtained for each $M_n$ sample. The plot reveals a $1/M_n$ dependence for $\tau_{on}$ between 55–110 kDa, in accordance with Eq. (2). On the other hand, $\tau_{off}$ remains constant at ~18 µs. This off-time is comparable to the triplet-state lifetime of P3HT found in deoxygenated environments [32, 33]. The "*off*"-state is completely quenched by oxygen [Fig. S7], a clear indication of triplet shelving. For $M_n < 45$ kDa (left side of the dashed line), $\tau_{on}$ deviates from the $1/M_n$ dependence because the acceptor is not fully formed in the intrachain self-aggregate, as illustrated in the inset [cf. PL spectra in Fig. 3]. The clear dependence of $\tau_{on}$ on $M_n$ supports the following conclusions: (i) the entire chain absorbs light (i.e. there is no "dark matter" [11]); (ii) EET occurs within the entire intrachain aggregate; (iii) the emitting chromophore can be shelved into the non-emissive triplet-state; and (iv) subsequent STA at the acceptor site quenches PL from the *entire* chain. These factors lead to the saturation of $\langle PL \rangle$ with increasing $M_n$ according to Eq. (4) for folded P3HT chains, as shown in Fig. 2(b). Quite surprisingly, even P3HT chains of $M_n = 110$ kDa behave as single-photon emitters. We conclude that, for folded P3HT chains, the only spectroscopic observable reporting the $M_n$ dependence is the excitation rate and therefore the average "*on*"-time.

Our results demonstrate the extraordinary light-harvesting properties of folded single P3HT chains, which are strongly dependent on the degree of ordering of the individual chain. On the one hand, ordering is required to design efficient solar cells [21]. On the other hand, order can also introduce losses through efficient STA, assuming that the dynamics of triplet formation observed here on single P3HT chains can be extrapolated to bulk heterojunction solar cells. Once a triplet is formed on the acceptor, additional singlets excited within the range of exciton migration are annihilated and cannot be converted to charges [34, 35]. This intrinsic loss-mechanism must contribute to the decrease of quantum efficiency in P3HT based organic solar cells with increasing excitation intensities, which has previously only been attributed to singlet-polaron interactions [36]. A simple estimate shows that the magnitude of STA observed on *single* P3HT chains can account for the observed bottleneck in solar cells. As described in Fig. S8, we consider the proportion of singlets lost due to STA at a given excitation intensity and singlet-triplet interaction distance, $D_{ST}$. For example, a loss of ~30 % is expected at intensities comparable to solar illumination for $D_{ST}$ = 30 nm. $D_{ST}$ depends on singlet and triplet exciton diffusion lengths. The singlet diffusion length in P3HT is reported to be in the range of tens of nanometers [35]. Triplet exciton migration is harder to quantify, but can even occur on the micron scale in organic semiconductor crystals [37]. Even in a percolative donor-acceptor blend with strong local order, $D_{ST}$ = 30 nm is likely an underestimate. STA should therefore be taken into account as a critical loss mechanism in organic solar cells, even though it is hard to separate from singlet-polaron quenching in the bulk [36]. A possible design route to prevent STA would be the inclusion of non-oxygenic triplet scavengers in the bulk heterojunction, selective introduction of disorder to lower $D_{ST}$, or identification of materials with weaker spin-orbit coupling and lower $k_{isc}$ than P3HT [32]. Organic solar cells bear remarkable similarity to photosynthesis in that energy conversion is fundamentally self-limiting [38], as observed in slowed plant growth under full-sun illumination.

In conclusion, a comprehensive monomolecular photophysical picture of the model organic-solar-cell compound P3HT emerges: the drop in fluorescence yield observed when going from unfolded to folded chains arises due to light absorption over the entire polymer, followed by efficient energy funneling to *one* emitting acceptor, which is subject to triplet blinking and PL saturation effects. The efficiency of plastic solar cells is therefore fundamentally limited by triplet accumulation. It will be intriguing to establish how well the model holds for well-ordered *multimolecular* interchain aggregates of similar $M_n$, which can be grown by solvent vapor annealing and offer a single-molecule route towards the bulk film [24, 39].

Acknowledgements

The authors are indebted to the ERC for funding through the Starting Grant MolMesON (305020). We thank Dr. T. Adachi for stimulating discussions.

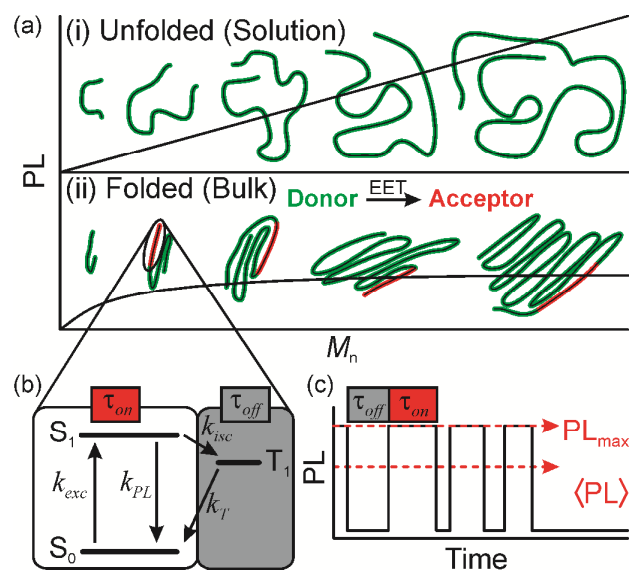

**Fig. 1.** (a) Photoluminescence (PL) intensity dependence on number-average molecular weight, $M_n$, for unfolded and folded CP chains. Efficient excitation energy transfer (EET) occurs for folded chains, so that only one chromophore emits (acceptor, red). (b) Level scheme of the acceptor chromophore. (c) The molecule is subject to triplet blinking with constant off-times, $\tau_{off}$, and $M_n$-dependent on-times, $\tau_{on}$, assuming the acceptor chromophore is populated by EET from the entire absorbing chain (donor, green).

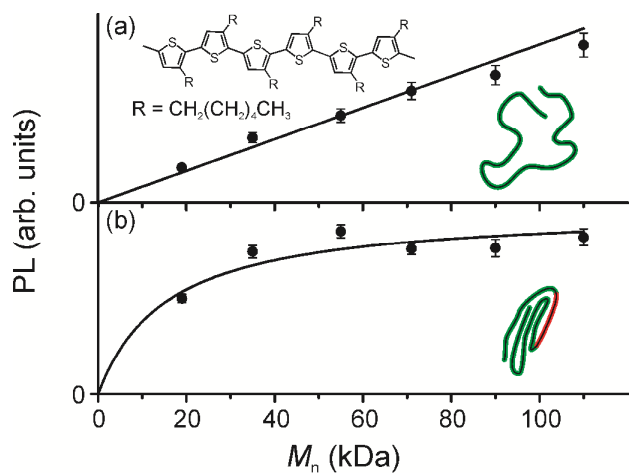

**Fig. 2.** (a) Perfectly unfolded and unaggregated regio-regular poly(3-hexylthiophene) (P3HT) chains (structure inset) only exist in solution. The average PL intensity was obtained by fluorescence correlation spectroscopy and plotted against $M_n$. (b) Folded chains are formed, on the single-chain level, by dispersion in a PMMA matrix. PL saturation with $M_n$ can be described by Eq. (4) (black curve). Error bars give the standard error.

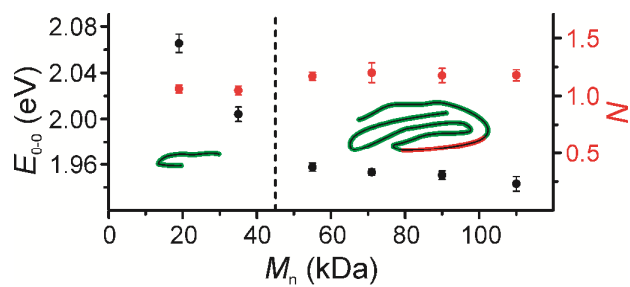

**Fig. 3.** The average number of independently emitting chromophores, $N$, in dependence of $M_n$ (red) for folded P3HT chains embedded in PMMA (see *Supporting Information* for explanation of error bars). The average transition energy of the single-chain PL peak, $E_{0-0}$, is shown in black. Folded chains display a constant $E_{0-0}$ of ~1.95 eV for $M_n > 45$ kDa. For shorter chains, the acceptor is not formed within all folded chains, as illustrated in the cartoon. Each data point gives an average over ~100 single chains. Error bars represent the standard error of the distribution (see *Supporting Information* for histograms).

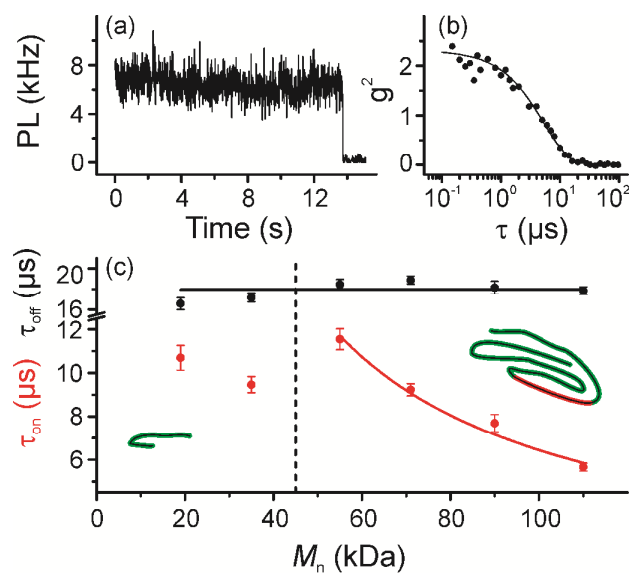

**Fig. 4.** (a) Typical PL transient for P3HT embedded in PMMA under nitrogen atmosphere, binned into 10 ms intervals. Single-step bleaching occurs after ~13 s. (b) Second-order cross-correlation, $g^2(\tau)$, obtained from the PL transient in (a); the line describes a single-exponential fit, which reveals fast blinking kinetics on the µs time scale. (c) For each $M_n$ ~100 transients were obtained, and the average on-times, $\tau_{on}$, (red dots) and off-times, $\tau_{off}$, (black dots) were extracted from the averaged $g^2(\tau)$ (see Fig. S6 and *Supporting Information* for detailed analysis). Error bars correspond to the fitting error. On-times for the two lowest-$M_n$ samples deviate from the $1/M_n$ dependence because the acceptor is not fully formed at this chain length, as illustrated in the cartoon.

Supporting Information for

**Singlet-triplet annihilation limits exciton yield in poly(3-hexylthiophene)**

Florian Steiner, Jan Vogelsang*, John M. Lupton

Institut für Experimentelle und Angewandte Physik, Universität Regensburg, Universitätsstrasse 31, 93053

Regensburg, Germany

**Materials and experimental apparatus.** Poly(3-hexylthiophene) (P3HT) with a regioregularity of 95.7 % was purchased from EMD Chemicals Inc. and further purified by GPC with a polystyrene standard to obtain 6 different $M_n$ samples with a low PDI (see Table S1). Poly(methylmethacrylate) (PMMA, $M_n$ = 46 kDa, PDI = 2.1) was purchased from Sigma Aldrich Co. Isolated chains of P3HT embedded in a ~50-nm-thick PMMA matrix were obtained by dynamically spin-coating at 2000 rpm from toluene after the following steps were executed: (i) Borosilicate glass cover slips were cleaned in a 2% Hellmanex III (Hellma Analytics) solution, followed by rinsing with MilliQ water. (ii) The glass cover slips were transferred into a UV-ozone cleaner (Novascan, PSD Pro Series UV) to bleach residual fluorescent molecules. (iii) P3HT was diluted in toluene to single-molecule concentration (~ $10^{-12} - 10^{-13}$ M) and mixed with a 1 wt% PMMA/toluene solution. To avoid photo-oxidation, the samples were prepared in a glove box and investigated in a microscope, described elsewhere in detail[1], using a home-built gas flow cell to purge with nitrogen gas. Briefly, a customized confocal scanning fluorescence microscope based on an Olympus IX71 was used. Excitation was carried out by either a fiber-coupled diode laser (PicoQuant, LDH-D-C-485) at 485 nm under cw excitation or with a white-light laser (NKT Koheras, SuperK Extreme) at 485 nm and 78 MHz for TCSPC experiments. The excitation light was passed through a clean-up filter (AHF Analysentechnik, z485/10), expanded and collimated via a lens system to a beam diameter of ~1 cm and coupled into an oil-immersion objective (Olympus, APON 60XOTIRF, NA = 1.49) through the back port of the microscope and a dichroic mirror (AHF Analysentechnik, z488RDC) for confocal excitation. The fluorescence signal passed a fluorescence filter (AHF Analysentechnik, RS488LP) and was split by a 50/50 beam splitter and detected by two avalanche photodiodes (APDs, PicoQuant, τ-SPAD-20). The excitation intensities were set to 50 W/cm² and 1 kW/cm² for confocal scanning and solution experiments, respectively. Time-tagged photon arrival was recorded by a TCSPC card (PicoQuant, Hydraharp 400) and further analysed by a home-written LabView program.

| Sample | 1 | 2 | 3 | 4 | 5 | 6 |
|---|---|---|---|---|---|---|
| $M_n$[a] (kDa) | 19 | 35 | 55 | 71 | 90 | 110 |
| PDI[b] | 1.56 | 1.2 | 1.11 | 1.09 | 1.09 | 1.09 |

**Table S1.** P3HT samples fractionated by GPC. [a] Number-average molecular weight purified by GPC against polystyrene standards, [b] polydispersity index.

**Fluorescence Correlation Spectroscopy (FCS).** FCS measurements were conducted on a home-built confocal fluorescence microscope as described in the experimental section. Briefly, a white-light laser (NKT Koheras, SuperK Extreme) set at 485 nm was used as the excitation source. The laser output was coupled into an oil-immersion objective (Olympus, APON 60XOTIRF, NA = 1.49). The fluorescence signal was split equally by a 50/50 beam splitter onto two APDs, recorded by a TCSPC card (PicoQuant, HydraHarp 400) and cross-correlated using a software package (PicoQuant, SymPhoTime 64). Cross-correlation curves from all P3HT samples, such as the representative curve presented in Figure S1 (black dots) were measured at an excitation intensity of 1 kW/cm² from well-dissolved P3HT/toluene solutions with a concentration of P3HT ranging from $10^{-10}$ to $10^{-11}$ M. Only the diffusion component was observed, and analyzed by fitting to a two-dimensional diffusion model (see Figure S1, red curve) [2]:

$$G(\tau) = \frac{1}{N} \cdot \left(1 + \frac{\tau}{\tau_D}\right)^{-1} \quad \text{(S1)}$$

where $N$ is the average number of the detected two-level emitters (i.e. fluorophores or chromophores) in the focal volume and $\tau_D$ corresponds to the average residence time of a single molecule in the focal volume. The molecular brightness was compared between the different number average molecular weights of the P3HT samples by dividing the average PL intensity by $N$.

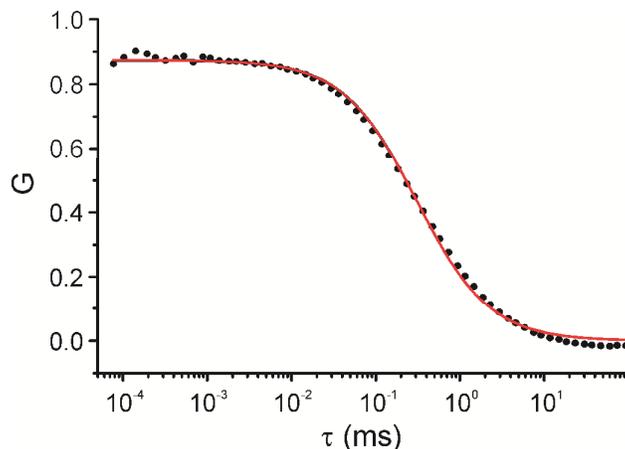

**Fig. S1.** FCS cross-correlation curve for sample 3 ($M_n$ = 55 kDa with PDI = 1.11, P3HT in toluene). The red curve corresponds to the fit according to equation (S1). $N$ was extracted from the fit and used to calculate the molecular brightness.

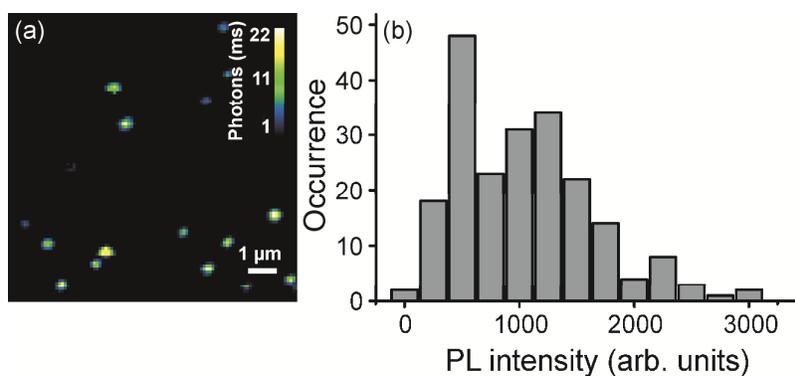

**Fig. S2.** a) Confocal fluorescence scan image of single P3HT chains embedded in a PMMA host matrix under dry nitrogen atmosphere at an excitation intensity of ~50 W/cm² and at an excitation wavelength of 485 nm. The spatial resolution was set to 100 nm/pixel with a time resolution of 8 ms/pixel. PL intensity histograms were obtained by extracting the PL intensity of the diffraction-limited spots by an automated spot-finding algorithm and subsequent background subtraction. Such a PL intensity histogram is shown in b) for sample 4 (P3HT with $M_n$ = 71 kDa and PDI = 1.09). A PL intensity histogram was obtained for each sample and the mean PL intensity with its corresponding standard error was calculated and plotted in Figure 1b and Figure S3.

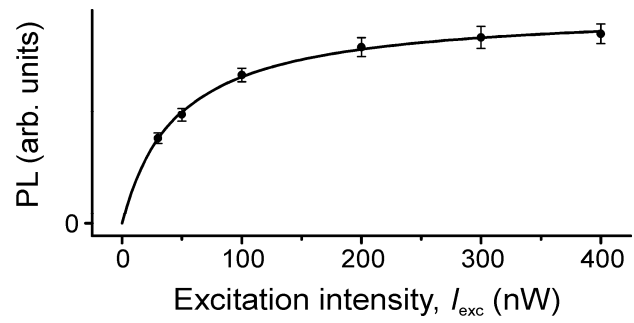

**Fig. S3.** PL saturation of folded P3HT chains with excitation intensity, $I_{exc}$, can be described by Eq. (4) (black curve). The error bars give the standard error.

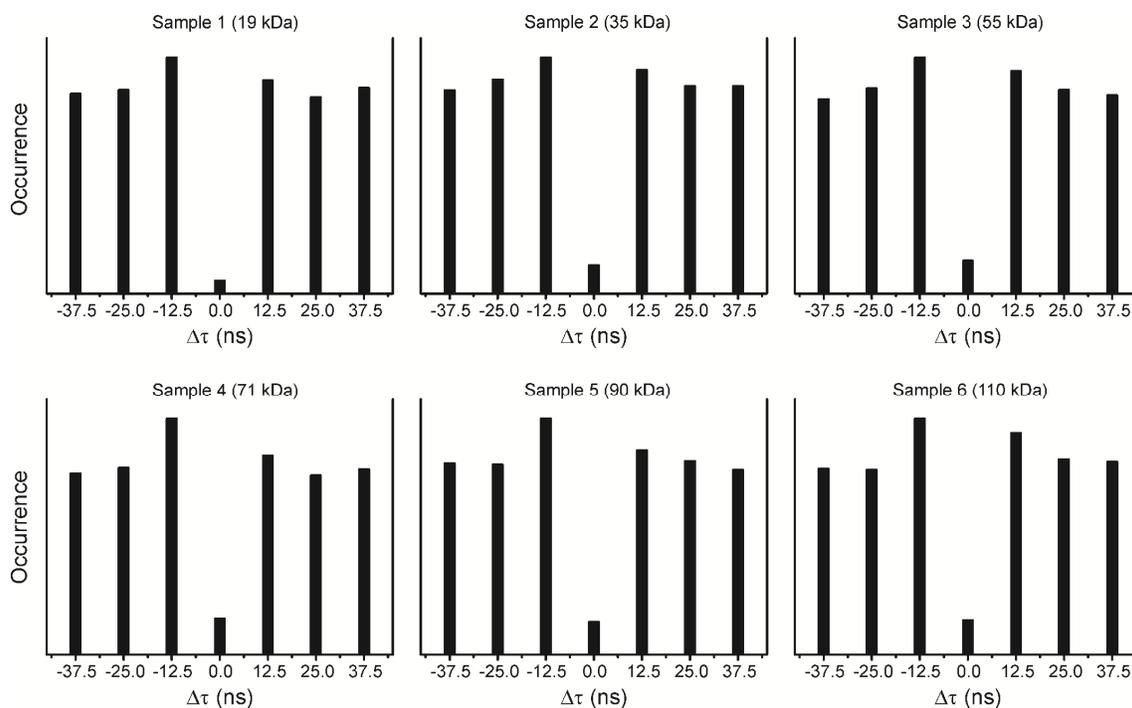

**Fig. S4.** Photon emission statistics obtained by the cross-correlation of two photodetectors in the emission pathway. The single P3HT chains embedded in PMMA were excited by laser pulses (12.5 ns period), controlling photon arrival times τ. A photon statistics histogram is presented for each $M_n$ sample, averaged over ~100 single P3HT chains. The number of independent emitters, $N$, was estimated from the ratio of the coincidence peak at time delay $\Delta\tau = 0$ to the lateral peaks at time delay $\Delta\tau \neq 0$ according to ref.[3], taking into account the signal-to-background correction. $N$ is plotted in Figure 2 as red dots. The error was estimated from the standard deviation of the 6 closest coincidence peaks. A dip in the coincidence rate of ~80% is observed for all $M_n$ samples because no more than a single photon is emitted from the polymer chain for each laser pulse.

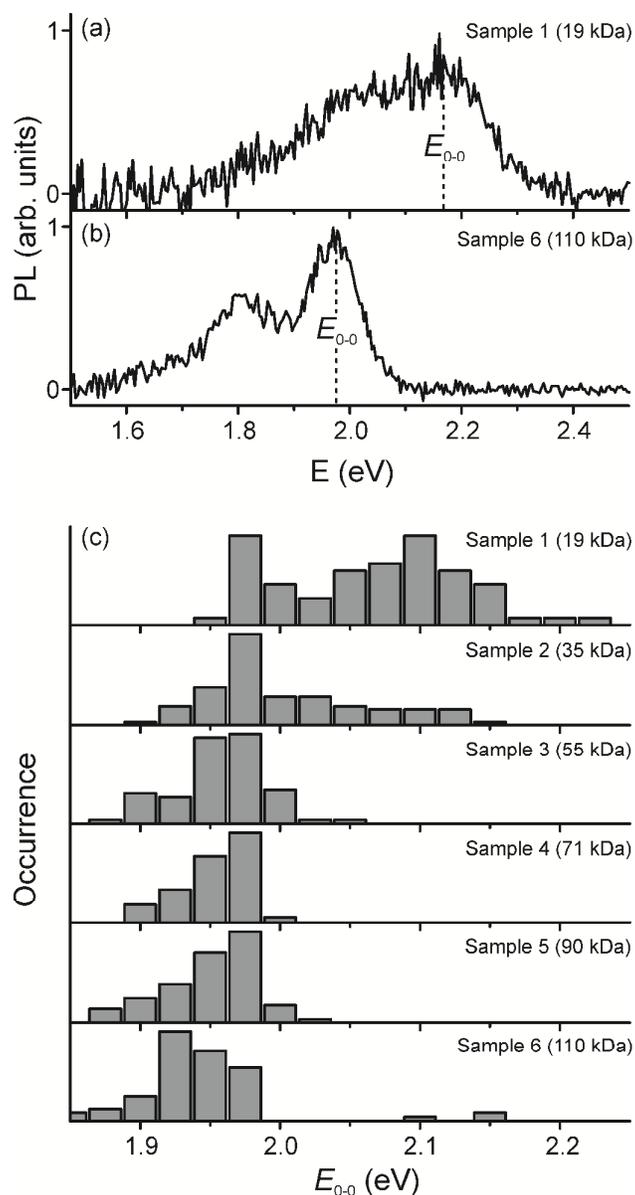

**Fig. S5.** a) Representative photoluminescence (PL) spectrum of a single P3HT chain of Sample 1 (19 kDa) embedded in a PMMA host-matrix. The 0-0 transition energy, $E_{0\text{-}0}$, was extracted for each spectrum by fitting the spectral shape to a Frank-Condon progression. b) Sample 6 (110 kDa), exhibiting a large red-shift of $E_{0\text{-}0}$. c) Histograms of the measured transition energies $E_{0\text{-}0}$ for each sample, as indicated. Each histogram represents approx. 100 single-molecule spectra. The average $E_{0\text{-}0}$ was obtained from these statistics and plotted in Figure 2 (black dots, including the standard error of the distribution).

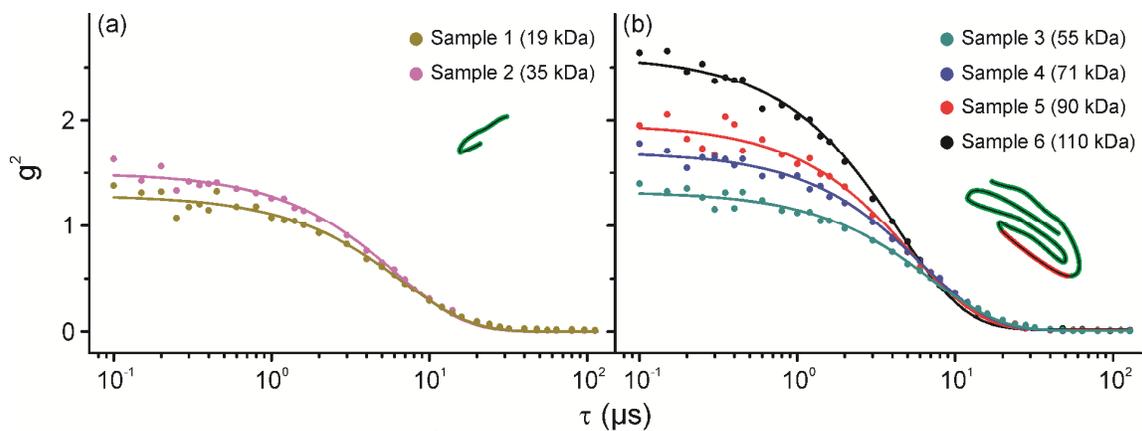

**Fig. S6.** Second-order cross-correlation, $g^2(\tau)$, curves obtained from single-molecule PL intensity transients. For each single-molecule PL transient (as shown in Figure 3a), a cross-correlation curve was computed, and approx. 100 curves were averaged for each sample. For clarity, the dataset is divided into the samples of incompletely folded chains (samples 1 and 2), in panel a); and the folded chains (samples 3 to 6), in b). These averaged cross-correlation curves can be very well described by a single-exponential decay (solid lines). See main text for a detailed interpretation of the dataset.

**Second-order cross-correlation analysis.** The analysis was conducted to extract the on- and off-times ($\tau_{on}$ and $\tau_{off}$) of single-molecule PL intensity transients as described in the literature[4].

First, the cross-correlation curves were computed with the following equation:

$$g^2(\tau) = \frac{\langle I_1(t) I_2(t+\tau) \rangle}{\langle I_1(t) \rangle \langle I_2(t) \rangle} - 1$$

with $I_1$ and $I_2$ corresponding to the measured intensities on the detection channel 1 and 2, respectively. The cross-correlation curves were fitted by using a single-exponential function

$$g^2(\tau) = A \cdot \exp\left(-\frac{\tau}{\tau_{ac}}\right)$$

as demonstrated in Figure S6. From this fit, the amplitude, $A$, and characteristic decay time, $\tau_{ac}$, were extracted. To evaluate the on- and off-times in the microsecond time regime corresponding to the formation of triplet-states, the on- and off-times were extracted from the background-corrected amplitude, $A_{bgcorr}$, and the characteristic decay time, $\tau_{ac}$, as follows:

$$\tau_{off} = \tau_{ac} \cdot (1 + A_{bgcorr})$$

$$\tau_{on} = \tau_{ac} \cdot \left(1 + \frac{1}{A_{bgcorr}}\right),$$

where $A_{bgcorr}$ is linked to the measured amplitude, $A$, as follows:

$$A_{bgcorr} = \left(\frac{S+B}{S}\right)^2 \cdot A$$

with S corresponding to the signal and B to the background of the transients.

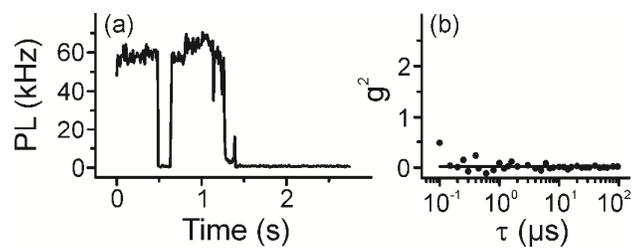

**Fig. S7.** a) Typical PL transient for P3HT embedded in PMMA under ambient conditions (i.e. with molecular oxygen present), binned into 10-ms intervals. A one-step bleaching event occurred after ~1.5 s. b) Second-order cross-correlation, $g^2(\tau)$, obtained from the PL transient in a); the figure features a flat cross-correlation curve: there are no blinking kinetics in the microsecond time regime since all triplets are quenched by $O_2$.

**Estimation of relative singlet exciton yield in bulk P3HT and implications for solar cells.** In the following, we want to estimate the relative singlet exciton yield due to singlet-triplet annihilation in the bulk of P3HT under low excitation intensities, $I_{exc}$, similar to sun light. The triplet yield of a single P3HT chromophore (approx. 42 %) is very high and therefore triplets can form easily within the P3HT bulk [5]. Once a triplet is formed, no additional triplets can be generated within a certain volume around this triplet, due to efficient singlet-triplet annihilation. This volume of P3HT material can then be described according to our 3-level system model (see Figure 1). The size of the volume, given by the singlet-triplet interaction distance, $D_{ST}$, determines the absorption cross-section, which is directly related to the degree of saturation of singlet excitons. The $S_1$ population at equilibrium can be calculated by setting the master equations for the 3-level system to zero:

$$0 = \frac{dS_0}{dt} = -k_{exc}S_0 + k_r S_1 + k_{nr} S_1 + k_T T_1$$

$$0 = \frac{dS_1}{dt} = +k_{exc}S_0 - k_r S_1 - k_{nr} S_1 - k_{isc} S_1 \quad \text{with} \quad S_0 + S_1 + T_1 = N = const.$$

$$0 = \frac{dT_1}{dt} = +k_{isc}S_1 - k_T T_1$$

From this follows that the population of $S_1$ equals:

$$S_1 = \left(\frac{k_r + k_{nr} + k_{isc}}{k_{exc}} + 1\right)^{-1} \cdot \left(1 + \left(\frac{k_r + k_{nr} + k_{isc}}{k_{exc}} + 1\right)^{-1} \cdot \frac{k_{isc}}{k_T}\right)^{-1} \cdot N$$

We compare this to the case of no triplets present, which equals $k_T$ going to infinity:

$$\lim_{k_T \to \infty} S_1 = \left(\frac{k_r + k_{nr} + k_{isc}}{k_{exc}} + 1\right)^{-1} \cdot N$$

Therefore the factor

$$\left(1 + \left(\frac{k_r + k_{nr} + k_{isc}}{k_{exc}} + 1\right)^{-1} \cdot \frac{k_{isc}}{k_T}\right)^{-1}$$

describes the relative singlet exciton yield, due to efficient singlet-triplet annihilation at a given excitation rate $k_{exc}$, which is given by:

$$k_{exc} = \sigma \cdot I_{exc}$$

with σ the absorption cross-section, which can be calculated for a given volume, $(D_{ST})^3$, of P3HT material by:

$$\sigma = \frac{\rho \cdot N_A}{M_w(C_{10}H_{14}S)} \cdot \sigma(C_{10}H_{14}S) \cdot (D_{ST})^3$$

Here, ρ is the bulk density of P3HT with 1.1 g·cm$^{-3}$ [6], $N_A$ is Avogadro's number, $M_w(C_{10}H_{14}S)$ is the molecular weight of one P3HT repeat unit with 166 g·mol$^{-1}$ and $\sigma(C_{10}H_{14}S)$ equals the absorption cross-section of one P3HT repeat unit with 1.002·10$^{-17}$ cm² at 485 nm wavelength [7]. With this, $k_{exc}$ becomes:

$$k_{exc}(D_{ST}, I_{exc}) = 4.008 \cdot 10^{-17} \cdot D_{ST}^3 \cdot I_{exc}$$

Together with the rates $k_r$ = (2ns)$^{-1}$ = 500 MHz, $k_{nr}$ = (1.5ns)$^{-1}$ = 666 MHz, $k_{isc}$ = (1.2 ns)$^{-1}$ = 833 MHz and $k_T$ = (17μs)$^{-1}$ = 59 kHz [7], the relative singlet exciton yield can be plotted versus the excitation intensity, $I_{exc}$, at 485 nm wavelength and the singlet-triplet interaction distance, $D_{ST}$, as shown in Figure S8.

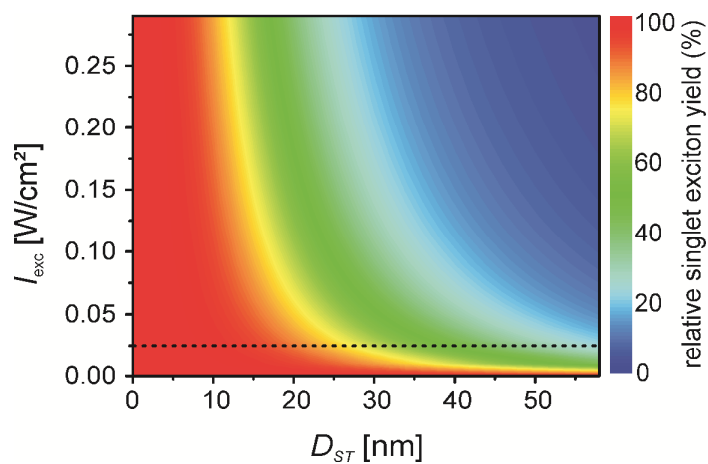

**Fig. S8.** Relative singlet exciton yield versus the singlet-triplet interaction distance, $D_{ST}$, and the excitation intensity, $I_{exc}$, at 485 nm wavelength. The dashed line corresponds approximately to the excitation regime of P3HT under AM1.5 conditions (0.1 W/cm² for the complete spectrum of the sun, therefore approximately 0.025 W/cm² including the absorption spectra of P3HT). Assuming a realistic $D_{ST}$ of ~30 nm the relative singlet exciton yield drops to ~70 %. In other words, roughly 30 % of singlet excitons are quenched in P3HT by a triplet under conditions relevant for organic solar cells.

We note that the intensity at 485 nm (the absorption maximum of P3HT) only provides an estimate of the overall excitation density under AM1.5 conditions; the spectrum of the sun needs to be matched to the absorption spectrum of P3HT.